\documentclass[11pt]{article}

\usepackage{amssymb}
\usepackage{graphicx}
\input epsf

\begin{document}

\begin{titlepage}
\begin{flushright}
CALT-68-2494
\end{flushright}

\begin{center}
{\Large\bf $ $ \\ $ $ \\
Supersymmetric null-surfaces}\\
\bigskip\bigskip\bigskip
{\large Andrei Mikhailov\footnote{e-mail: andrei@theory.caltech.edu}}
\\
\bigskip\bigskip
{\it California Institute of Technology 452-48,
Pasadena CA 91125 \\
\bigskip
and\\
\bigskip
Institute for Theoretical and 
Experimental Physics, \\
117259, Bol. Cheremushkinskaya, 25, 
Moscow, Russia}\\

\vskip 1cm
\end{center}

\begin{abstract}
Single trace operators with the large R-charge in supersymmetric
Yang-Mills theory correspond 
to the null-surfaces in $AdS_5\times S^5$. We argue that the
moduli space of the null-surfaces is the space of contours
in the super-Grassmanian parametrizing the
complex $(2|2)$-dimensional subspaces of the complex
$(4|4)$-dimensional space. The odd coordinates on this
super-Grassmanian correspond to the fermionic degrees
of freedom of the superstring. 
\end{abstract}

\end{titlepage}

\section{Introduction.}
In the AdS/CFT correspondence the single-trace
operators of the large $N$ supersymmetric Yang-Mills theory correspond 
to the single string states in the Type IIB theory on $AdS_5\times S^5$.
The single-trace operators have a form of the trace of the product of $L$
elementary fields of the Yang-Mills theory.
There is a certain subclass of these operators 
for which the corresponding string 
can be described semiclassically 
\cite{FT02,Tseytlin,Russo,TseytlinReview}.
 For  an operator
to correspond to a classical dual string, one has to take 
$L\to\infty$. Also, in taking this limit, one has to arrange
the elementary fields under the trace in such a way that
the operator is ``locally half-BPS''\cite{MMT}. 

Let us explain what ``locally half-BPS'' means. 
 The $N=4$ theory has six scalar fields $\Phi_1\ldots, \Phi_6$. 
Let us consider a complex combination $Z=\Phi_5+i\Phi_6$. In the limit 
$L\to\infty$ we have to require that each elementary field under the 
trace is of the form $\phi_k=g_k.Z$ where $g_k$ is some element
of the superconformal group $SU(2,2|4)$. In other words, each elementary 
field is in the superconformal orbit of $Z$.  Moreover 
we should have $g_{k+1}=g_k+O(1/L)$. Therefore instead of
the discreet ``chain'' of the group elements $g_1,\ldots, g_L$
we have a continuous contour in the group manifold $g(\sigma)$,
where $\sigma=2\pi k/L$. 
In this ``continuous limit''
the anomalous dimension of the Yang-Mills operator 
becomes of the order $\lambda\over L$ plus the higher
order corrections which are the series in $\lambda\over L^2$.
Moreover, the renormgroup flow defines a classical dynamical
system on the space of contours $g(\sigma)$ 
(see \cite{Kruczenski} and references therein).
More precisely, $g(\sigma)$ takes values not in the
group manifold $PSU(2,2|4)$ itself but rather in the coset
space which is $PSU(2,2|4)$ modulo the subgroup which
acts on $Z$ as a phase rotation.
Therefore the renormgroup flow in the field theory defines
a classical dynamical system on the space of
loops $g(\sigma)$ taking values in the supercoset
$Gr(2|2,4|4)=U(2,2|4)/(U(2|2)\times U(2|2))$. 

This supercoset has dimension $(16|16)$, sixteen even and sixteen 
odd coordinates. Therefore the ``continuous''
Yang-Mills operators are described
by $16$ even and $16$ odd functions of one real variable.
But $16$ even and $16$ odd functions also parametrize the
phase space of the classical Type IIB superstring
in $AdS_5\times S^5$. Therefore, it is natural to
conjecture that the classical dynamical system on 
the space of the locally half-BPS operators defined by the
renormgroup flow is equivalent to the classical
worldsheet dynamics of the Type IIB superstring.

Unfortunately we do not know any independent prescription
which would tell us which string worldsheet corresponds
to a given Yang-Mills operator. But the conjecture
that the string sigma-model is equivalent to the
classical renormgroup flow is nontrivial even without
such a prescription. Indeed, the equivalence of two
dynamical systems is already a nontrivial statement.
It was partially verified at the two loop
level (in three different ways!) in \cite{AS,KMMZ,KRT}.
 
Let us briefly review what happens at the one loop 
level, following \cite{SS,SE} 
(see also \cite{HernandezLopez,StefanskiTseytlin} for a different
approach).
Consider the string worldsheet corresponding to
a given Yang-Mills single trace operator composed of $L$ 
elementary fields. The shape of the worldsheet depends on
the coupling constant $\lambda$.
When ${\lambda\over L^2}\to 0$ the worldsheet
degenerates and becomes a null-surface. Moreover,
this null-surface comes with a parametrization of the
light rays. Therefore in the ``continuous'' limit the 
single trace operators correspond to the parametrized
null surfaces. It turns out that the string worldsheet
theory defines the structure of a Hamiltonian system
on the space of parametrized null-surfaces. The definition
of this Hamiltonian system goes as follows. 
Pick a parametrized null-surface. Consider a family
of extremal surfaces depending on the parameter $\epsilon$,
such that: 1) the limit when $\epsilon\to 0$ is our
null-surface and 2) the density of the conserved charges
on the worldsheet in the limit $\epsilon\to 0$ 
is proportional to ${1\over\epsilon}d\sigma$ where
$\sigma$ is the parametrizing function (see Section 3.3
of \cite{SE} for the precise formula). There are infinitely many such
families, but for the purpose of our definition
we can pick any one, satisfying these two
properties. The deviation of this extremal surface from our
null-surface is locally of the order $\epsilon^2$. But if we
follow its evolution in the global AdS time, the deviation
will accumulate. After the time interval of the order
$\Delta T\sim {1\over\epsilon^2}$, the deviation becomes of the order
one, and our extremal surface will locally approximate
(with the accuracy $\sim\epsilon^2$) another null-surface. 
This determines the evolution on the space of 
null-surfaces. 

An important point is that this definition
does not depend on the choice of the family converging
to the null-surface. If we pick two different worldsheets
approximating the same parametrized null-surface, they will ``oscillate''
around each other, but the deviation between them will not
accumulate in time. Therefore, different approximating
surfaces determine the same ``slow evolution'' on the
space of null-surfaces. 

The space of parametrized null-surfaces
is identified with the space of pairs of functions $Y:\;S^1\to {\bf C}^{2+4}$,
$Z:\; S^1\to {\bf C}^{6}$ such that $|Y(\sigma)|^2=|Z(\sigma)|^2=2$
and $(Y(\sigma),Y(\sigma))=(Z(\sigma),Z(\sigma))=0$ and 
$(\overline{Y},\partial_{\sigma}Y)=(\overline{Z},\partial_{\sigma}Z)$.
The one-loop anomalous dimension corresponds to the Hamiltonian
of the slow evolution:
\begin{equation}\label{Delta}
\Delta={1\over 16\pi^2}{\lambda\over (L/2\pi)}
\int_0^{2\pi} d\sigma \left(
(\partial_{\sigma}\overline{Z},\partial_{\sigma}Z)-
(\partial_{\sigma}\overline{Y},\partial_{\sigma}Y)\right)
\end{equation} 
This space is a $U(1)$-bundle over a submanifold of the loop space
of the product of two Grassmanians:
\begin{equation}\label{BosonicCoset}
{SO(2,4)\over SO(2)\times SO(4)}\times {SO(6)\over SO(2)\times SO(4)}
\end{equation}
It consists of the loops satisfying certain integrality condition 
which corresponds to the cyclic invariance of the trace.

In this paper we will argue
that the fermionic degrees of freedom on the worldsheet 
parametrize in the ultrarelativistic limit the
odd directions of the supercoset space. Just as the fast-moving
bosonic string corresponds to a contour in the product
of two Grassmanians (\ref{BosonicCoset}) the
superstring defines a contour in 
$${U(2,2|4)\over U(2|2)\times U(2|2)}$$
Turning on the fermionic degrees of freedom of the null-surface
corresponds on the field theory side to considering
operators with insertions of the fermions and the field strength.

Dynamical systems on supersymmetric coset spaces were
studied in the recent papers \cite{IvanovMezincescuTownsend}.

The null-surface perturbation theory was previously studied in a closely
related context in \cite{dVGN}.

\section{Single trace operators with large spin.}
The single trace operators are the operators of the form
$\mbox{tr}\; \phi_1\phi_2\cdots\phi_n $ where $\phi_1,\ldots,\phi_n$
are the fundamental fields.
The one loop anomalous dimension for all the single trace operators 
in the ${\cal N}=4$ Yang-Mills theory was computed 
in \cite{MinahanZarembo,SSC}.
For the one loop computation, each fundamental field under the trace 
can be considered a free field, and therefore it transforms in the 
singleton representation of the superconformal group.
The one loop anomalous dimension corresponds to the hermitean operator 
(interaction Hamiltonian) acting on the trace of the product 
of the free fields. 
It was shown in \cite{SSC} that this operator commutes with
the generators of the superconformal group. 
In the planar limit the interaction is a sum of the
pairwise interactions of all the 
fundamental fields under the trace which are
next to each other in the product. In other words, the one loop anomalous
dimension of the operator $\mbox{tr}\;\phi_1\phi_2\cdots\phi_n$
is given by the sum of the diagramms involving
$\phi_1$ and $\phi_2$, diagramms involving $\phi_2$ and $\phi_3$
and so on. The one loop interaction preserves the number of the
fundamental fields under the trace.

\subsection{Coherent states.}
The ``continuous'' approach to the computation of the anomalous dimension
of the single trace operator was proposed in \cite{Kruczenski}.
This approximation is useful when the number of the fields
under the trace is very large. This approach (as we understand it)
relies on the existence of the special set of vectors in the
singleton representation. This set of vectors is obtained in the
following way. Consider the free ${\cal N}=4$ theory 
on ${\bf R}\times S^3$. The vacuum is conformally invariant.
Let us act on the vacuum by the creation operator of the free boson 
$Z(x)=\phi_5(x)+i\phi_6(x)$ integrated over $S^3$.
We will get a state which we will call $\psi_{\bf 1}$. This state
is not invariant under $PSU(2,2|4)$. For any group element 
$g\in PSU(2,2|4)$ we will denote $\psi_g=g.\psi_{\bf 1}$.
We will call the states of the form $g.\psi_{\bf 1}$ 
{\em coherent}.

It is important to identify the subgroup $H\subset PSU(2,2|4)$ 
which acts on $\psi_{\bf 1}$ by rotating its phase. 
Let us first discuss the bosonic part of $H$. The bosonic part
of $PSU(2,2|4)$ is\footnote{The global structure of the
superconformal group is not very important here because
the center of the superconformal group is a subgroup of $H$. 
We will consider
the global issues in Section \ref{sec:Fast}.}
$SU(2,2)\times SU(4)$.
And the bosonic part of $H$ consists of the shift of time
and the isometries of $S^3$, which together form 
$S(U(2)\times U(2))=(SU(2)\times SU(2)\times U(1))/{\bf Z}_2$,
 plus the subgroup 
$S(U(2)\times U(2))$ 
of the R-symmetry $SU(4)$ which
preserves the direction of $Z(x)$ in the isotopic space:
\begin{equation}
	H_{rd}=S(U(2)\times U(2))\times S(U(2)\times U(2))
\end{equation}
This is (modulo the global structure) the even part of the supergroup:
\begin{equation}
	H=PS(U(2|2)\times U(2|2))
\end{equation}
This supergroup can be understood as the central extension
of $PSU(2|2)\times PSU(2|2)$ by the $U(1)$ which we call
$U(1)_c$, and another $U(1)$ generated by the ``grading element $d$''
entering as a semidirect product:
\begin{equation}\label{CentralExtension}
	PS(U(2|2)\times U(2|2))=U(1)_d \ltimes 
	\left(PSU(2|2)\times PSU(2|2)\right){\ltimes} U(1)_c 
\end{equation}
Here $\ltimes$ denotes the semidirect product\footnote{In the original
version of this paper we wrote incorrectly that the two factors
$U(1)$ enter through the direct product. In fact this is
a semi-direct product. The most economical notation for $H$ is
$PS(U(2|2)\times U(2|2))$. We realized the mistake studying
the recent works \cite{ArutyunovFrolov0411,BKSZ,AAT}. 
If there were two free factors $U(1)$ then the classical
superstring in $AdS_5\times S^5$ would have two series of local
conserved charges; but in fact there is only one from $U(1)_d$.}; 
if $M$ and $N$ are
two groups then $G=M\ltimes N$ means that $N$ is a normal
subgroup in $G$ and $G/N=M$.
The purely bosonic part of (\ref{CentralExtension}) contains
both $U(1)_d$ and $U(1)_c$ as free factors. But once we 
turn on the odd coordinates of the supergroup, 
$U(1)_d$ does not commute with the
fermionic generators of $(PSU(2|2)\times PSU(2|2))\ltimes U(1)_c$ 
and also $PSU(2|2)\times PSU(2|2)\subset U(1)_d \ltimes 
	\left(PSU(2|2)\times PSU(2|2)\right){\ltimes} U(1)_c$
	is not a subgroup because the anticommutator 
	of two odd elements in $PSU(2|2)$ generally speaking has
	a component in $U(1)_c$.
In fact $H$ can be defined as the centralizer of $U(1)_c$ in
$PSU(2,2|4)$, which makes the contact with \cite{ArutyunovFrolov0411,BKSZ,AAT}.
In the plane wave language $U(1)_c$ corresponds to 
the lightlike Killing vector field $\partial \over\partial x^-$,
and $U(1)_d$ corresponds to the other Killing vector field
$\partial\over\partial x^+$ which is not lightlike. 
The coordinate $x^+$ is considered a time coordinate in the plane
wave language. The subgroup $U(1)_d$ does not commute with
the fermionic part of $(PSU(2|2)\times PSU(2|2)) \ltimes U(1)_c$
because the generators of the supersymmetry in the plane
wave background are time-dependent 
\cite{BlauFigueroaOFarrillHullPapadopoulos,Metsaev,BMN}.

Therefore the coherent states are parametrized
by the points of the coset space 
\begin{equation}\label{Grassmanian}
Gr(2|2,4|4)={PSU(2,2|4)\over PS(U(2|2)\times U(2|2))}=
{U(2,2|4)\over U(2|2)\times U(2|2)}
\end{equation}
These states generate the singleton representation. 
It is then conjectured that in the limit of the large number
of fields the ``semiclassical'' states are the decomposable tensors
of the form 
\begin{equation}
\psi_{g_1}\otimes\psi_{g_2}\otimes\cdots\otimes\psi_{g_n}
\end{equation}
where the difference between $g_k$ and $g_{k+1}$ is of the
order $1\over n$. In the continuum limit $n\to\infty$,
the number of the site $k$ becomes a continuous parameter
$\sigma=k/n$, and the evolution of the state is
approximated by the classical evolution of the
contour $g(\sigma)$. The classical Hamiltonian is the matrix
element
\begin{equation}\label{ClassicalHamiltonian}
H_{cl}[g(\sigma)]=
(\overline{\psi}_{g_1}\otimes\overline{\psi}_{g_2}\otimes
\cdots\otimes\overline{\psi}_{g_n},\; H_{int}\;\;
\psi_{g_1}\otimes\psi_{g_2}\otimes\cdots\otimes\psi_{g_n})
\end{equation}
The symplectic structure is $\Omega=\int d\sigma\;\Omega(\sigma)$
where $\Omega(\sigma)$ is the differential of the one-form
\begin{equation}\label{OneForm}
\alpha(\sigma)=(\overline{\psi}_{g(\sigma)},d\psi_{g(\sigma)})
\end{equation}
This data defines a dynamical system on the space of contours
in the super-Grassmanian (\ref{Grassmanian}).
The interaction Hamiltonian of \cite{SSC} involves only
the pairs of neighbors in the product. Therefore
the continuous Hamiltonian should be a local functional
of the contour. Moreover, one can see that it contains
not more than two derivatives $\partial_{\sigma}$.
Therefore the value of the Hamiltonian on the 
contour should be given by the value of the classical
action of the free particle on the super-Grassmanian
which has this contour as a trajectory\footnote{The contour
can be an arbitrary trajectory, not necessarily
satisfying the equations of motion.}.

It is useful to write down these coherent states more
explicitly. The space of one-particle states
of the free scalar field theory can be identified with the space of the
positive-frequency solutions of the free field equations.
We will describe a family of positive-frequency solutions
parametrized by the points of the coset 
${SO(2,4)\over SO(2)\times SO(4)}\times 
 {SO(6)\over SO(2)\times SO(4)}$.
This coset is parametrized by two
complex lightlike vectors $Y$ and $Z$, 
$(\overline{Y},Y)=(\overline{Z},Z)=2$, $(Y,Y)=(Z,Z)=0$, modulo
independent phase rotations of $Y$ and $Z$.
The manifold of the complex lightlike vectors $Y$ in
${\bf C}^{2+4}$ consists of two connected
components. Those $Y$ which can be rotated by $SO(2,4)$ 
to $(1,-i,0,0,0,0)$ belong to the first component, and those which
can be rotated to $(1,i,0,0,0,0)$ belong to the second component.
We need only the first component.
Given $Y$ and $Z$, let us consider the positive frequency
wave of the field $Z_1\Phi_1+\ldots+Z_6\Phi_6$ parametrized by $Y$:
\begin{eqnarray}
&&f_{[Y]}(\tau,{\bf n})= {1\over Y_{-1}\cos\tau +Y_{0}\sin\tau
-({\bf Y},{\bf n})}\\[5pt]
&&\left({\partial^2\over\partial\tau^2}-
{\partial^2\over\partial{\bf n}^2}\right)f_{[Y]}=
-f_{[Y]}\nonumber
\end{eqnarray}
Here   $(\tau,{\bf n})$ are the
coordinates on ${\bf R}\times S^3$, $\tau$ is 
parametrizing ${\bf R}$ 
and ${\bf n}=(n_1,n_2,n_3,n_4)$ is the unit vector parametrizing 
$S^3$. This defines a state $\psi_{[Y],[Z]}$.

The super-Grassmanian (\ref{Grassmanian}) is the (4,2,2) analytic
superspace of \cite{GIKOS,LN,HHHW}. The bosonic coset space
${SO(2,4)\over SO(2)\times SO(4)}\simeq 
 {SU(2,2)\over S(U(2)\times U(2))}$ 
was discussed in the context of the AdS/CFT correspondence
in \cite{Gibbons}, where it was identified as the moduli
space of timelike geodesics in $AdS_5$.
 This coset space is 
the future tube of the Minkowski spacetime
(see \cite{Gibbons} and references therein, and 
\cite{MP,VladimirovSergeev} for the mathematical background). 
The motion
of a particle in AdS space was also studied in
\cite{Gradechi}. An interesting technique
for obtaining the singleton representation
from dynamics in higher dimensional spaces
was developed in \cite{Twotimes}; in the next section
we will use an approach related to the ideas of 
\cite{Twotimes} to study spinors in AdS space. 
The general theory of the coherent states of the
type discussed in this section was developed in \cite{MP}.

The singleton representation of $SO(2,4)$ is not square
integrable (see \cite{Starinets} for a recent discussion of
this fact, and references therein). Therefore our system
of coherent states does not resolve the identity.

\subsection{Anomalous dimension.}
The coherent states of the spin chain generally speaking
do not have a definite energy in the $\lambda=0$ theory.
In fact the corresponding operators are superpositions of
operators with different engineering dimensions.
In this situation we can define the one-loop anomalous 
dimension in the following way. 
The superconformal group is actually 
$\widetilde{PSU}(2,2|4)$ --- the universal covering of 
$PSU(2,2|4)$.
The universal covering has a center $C={\bf Z}$.
Let $c$ denotes the generator of the center.
In the free field theory $c=1$, but it acts nontrivially 
in the interacting theory:
\begin{equation}
	c=e^{i \Delta}
\end{equation}
In perturbation theory
$\Delta$ is expanded in powers of $\lambda$. It starts
with the term linear in $\lambda$, which is the
one-loop anomalous dimension. It is obvious from this
definition that the one-loop anomalous dimension commutes
with the superconformal group.

At the one-loop level
the action of the center is the sum of the contributions
of the pairwise interactions of the nearest neighbors
of the parton chain. Each pairwise interaction separately
commutes with the superconformal group. Therefore 
it is enough to consider the case when the pair of
nearest neighbors is:
\begin{equation}
	\ldots\otimes Z(0)\otimes 
	(Z(0)+a {\cal O}_1+ a^2{\cal O}_2+\ldots)\otimes\ldots
\end{equation}
Here $a$ is the lattice spacing and  ${\cal O}_n$ are some
combinations of elementary fields. For the continuous
operators
\begin{equation}\label{DeltaZ}
{\cal O}_1=\alpha_I\Phi^I(0)+\beta^{\mu}\partial_{\mu}Z(0)
\end{equation}
where $\alpha_I$ ($I=1,2,3,4$) and
$\beta^{\mu}$ are some complex coefficients. 
We need to compute the expectation value of the
interaction Hamiltonian in the coherent state:
$$
\langle H\rangle=	\left(\overline{Z}(0)\otimes 
	(\overline{Z}(0)+a\overline{\cal O}_1+\ldots),
	H\; Z(0)\otimes (Z(0)+a{\cal O}_1+\ldots)\right)
$$	
Notice that $H.Z(0)\otimes Z(0)=0$ because 
$Z(0)\otimes Z(0)$ is the vacuum (``the BMN vacuum'').
Therefore 
\begin{equation}\label{PairExpectation}
\langle H\rangle =
a^2(\overline{Z}(0)\otimes \overline{\cal O}_1,\;
	H\;\; Z(0)\otimes {\cal O}_1)+o(a^2)
\end{equation}
All that we need in the continuum limit is the leading term.
The action of $H$ was computed in \cite{SSC}. We will now
briefly review some results of \cite{SSC}.
The tensor product of two supersingletons $V_F$
of $PSU(2,2|4)$
is a reducible representation. It decomposes
into the direct sum of irreducible representations
of $PSU(2,2|4)$:
\begin{equation}
	V_F\otimes V_F=\bigoplus\limits_{j=0}^{\infty} V_j
\end{equation}
The action of $H$ is the sum of 
the projectors $P_j$ on $V_j$, with the
coefficients depending on $j$. The coefficient
of $P_0$ is zero, and the coefficient
of $P_1$ is $\lambda\over 4\pi^2$.
The BMN vacuum $Z(0)\otimes Z(0)$ belongs to $V_0$. 
What can we say about $Z(0)\otimes {\cal O}_1$ 
where ${\cal O}_1$ is given by (\ref{DeltaZ})?
One can see that the symmetric part
$Z(0)\otimes {\cal O}_1 + {\cal O}_1\otimes Z(0)$
belongs to $V_0$ and the antisymmetric part
$Z(0)\otimes {\cal O}_1 - {\cal O}_1\otimes Z(0)$
belongs to $V_1$. This means that:
\begin{equation}
	(\overline{\Psi}, H.\Psi)=(\overline{\Psi},
	{\lambda\over 8\pi^2}
\sum\limits_{l=1}^L (1-P_{l,l+1})\Psi)+\ldots
\end{equation}
where dots denote terms subleading in the continuous limit.
To compute the right hand side, we need the scalar product
$(\overline{\psi}_{[Y],[Z]},\psi_{[Y'],[Z']})=
{(\overline{Z},Z')\over (\overline{Y},Y')}$. 
Therefore the one-form (\ref{OneForm}) is 
$(\overline{Y},dY)-(\overline{Z},dZ)$, and
the classical Hamiltonian
$(\overline{\Psi}, H.\Psi)$ is given by (\ref{Delta}).

\section{Fast moving superstrings.}\label{sec:Fast}
\subsection{Anomalous dimension as a deck transformation.}
AdS space is the universal covering space of the
hyperboloid. The center of $\widetilde{PSU}(2,2|4)$
acts as a deck transformation exchanging the sheets.
We can visualize the action of this deck transformation
on the string phase space in the following way.
Let us replace $AdS_5$, which is the covering
space of the hyperboloid, by the hyperboloid
itself $AdS_5/{\bf Z}$. Let us formally consider
the string as living on $(AdS_5/{\bf Z})\times S^5$.
Let us pick a point $x$ on the string worldsheet $\Sigma$.
Consider a neighborhood of $x$ in 
$(AdS_5/{\bf Z})\times S^5$ 
which is simply connected.
For example, we can pick as such a neighborhood a set
of points  which are within the distance
$R/2$ from $x$, where $R$ is the radius of $AdS_5$.
Let $B$ denote such a neighborhood. Consider
the part of the string worldsheet which is inside
$B$ (that is, $B\cap \Sigma$). One can see that
$B\cap \Sigma$ consists of several 
sheets, which can be enumerated. 
These sheets are two-dimensional, so we can think of
them as cards; $B\cap \Sigma$ is then a deck of 
cards. 
Let $x$ belong to 
the sheet number $n$, then we can draw a path
on $\Sigma$ starting at $x$, winding once on 
the noncontractible cycle in $AdS_5/{\bf Z}$ and
then ending on the sheet number $n+1$. 
Let $\Sigma_n$ denote the sheet number $n$.
The coordinate
distance between $\Sigma_n$ and $\Sigma_{n+1}$
is of the order $\epsilon^2$.
The deck transformation maps $\Sigma_n$ to $\Sigma_{n+1}$,
$\Sigma_{n+1}$ to $\Sigma_{n+2}$ and so on.
This determines the action of the discreet group 
${\bf Z}$ on the string phase space, which should
be identified with the action of the center of the
superconformal group on the field theory side. 

In this paper we consider only the one-loop approximation.
Therefore we need the action of $\bf Z$ to the
order $\epsilon^2$. At this order the 
deck transformation can be interpreted
as the slow evolution.    
Indeed, let us replace $n$ with the continuous
parameter $t=n\epsilon^2$.
For every $n$ the corresponding sheet $\Sigma_n$ is close
to some null-surface which we denote $\Sigma(0)^{(t)}$;
this null-surface is defined for each $t$; there is
an ambiguity in the definition of $\Sigma(0)^{(t)}$ but it
is of the order $\epsilon^2$. Therefore, in the limit
$\epsilon^2\to 0$ the deck transformations define
a one-parameter family of transformations of the
null-surfaces. This slow evolution
of the null surfaces was studied in \cite{SE}
but only in the bosonic sector.
In this section we will turn on the fermionic
degrees of freedom.

\subsection{Supersymmetric null surfaces.}
The worldsheet theory for the
superstring in $AdS_5\times S^5$ can be formulated as a sigma-model
with the target space the supercoset
\begin{equation}
M={PSU(2,2|4)\over SO(1,4)\times SO(5)}
\end{equation}
In the ultrarelativistic limit the string worldsheet becomes
a parametrized null-surface. The parametrized null-surfaces correspond
to the loops in the coset space 
$$
{SO(2,4)\over SO(2)\times SO(4)}\times
{SO(6)\over SO(2)\times SO(4)}
\simeq 
{SU(2,2)\over S(U(2)\times U(2))}\times {SU(4)\over S(U(2)\times U(2))}
$$
This is the bosonic part of the story.
Let us study the effect of the fermionic degrees of freedom.

We will apply the Green-Schwarz approach for the string in $AdS_5\times S^5$
developed in \cite{MetsaevTseytlin}, but  using a different
representation of the gamma-matrices. 
Following \cite{Bar} we will represent the 
spinors in the space $AdS_5\times S^5$ as restrictions of spinors
in the flat space ${\bf R}^{2+10}$. 
We prefer this representation, because
the covariantly constant spinors on
$AdS_5\times S^5$ correspond in this picture to the constant spinors in
the flat space.
This also agrees with the philosophy of \cite{Twotimes}.
Consider the embeddings 
$AdS_5\subset {\bf R}^{2+4}$ with coordinates $X_{-1},X_0,\ldots,X_4$
and $S^5\subset {\bf R}^6$ with coordinates $X_5,\ldots,X_{10}$.
Consider the twelve-dimensional chiral spinors $\Psi$ of $SO(2,10)$.
These twelve-dimensional spinors are sections of the
spinor bundle over ${\bf R}^{2+10}$, which is a trivial
bundle (the product ${\bf R}^{2+10}\times {\bf C}^{32}$).
Let us restrict this bundle to $AdS_5\times S^5\subset {\bf R}^{2+10}$.
This restriction would be a trivial bundle 
$AdS_5\times S^5\times {\bf C}^{32}$.
It turns out that the spinor bundle on $AdS_5\times S^5$
can be realized as a subbundle of this bundle. 
This subbundle is the image of the projector
${1\over 2}(1+\Gamma^{A}\Gamma^{S})$ where $\Gamma^A$ is the 
$\Gamma$-matrix corresponding to the vector of the length
square $-1$ orthogonal to $AdS_5$ in ${\bf R}^{2+4}$ 
and $\Gamma^S$ corresponds
to the unit vector orthogonal to $S^5$ in ${\bf R}^6$.
Therefore 
a section of the spinor bundle on $AdS_5\times S^5$
can always be represented in the form: 
\begin{equation}\label{Projector}
\psi={1\over 2}
\left(1+\Gamma^{A}\Gamma^{S}\right)\Psi_{++}
\end{equation}
where $\Psi_{++}$ satisfies:
\begin{eqnarray}
\Gamma_{-1}\Gamma_0\cdots\Gamma_4\Psi_{++}=i\Psi_{++}\nonumber\\[-5pt]
\label{PP}\\[-5pt]
\Gamma_5\cdots\Gamma_{10}\Psi_{++}=i\Psi_{++}\nonumber
\end{eqnarray}
This condition means that $\Psi_{++}$ is the product of
the positive chirality spinor of $SO(2,4)$ and the positive
chirality spinor of $SO(6)$.
We will choose the $\Gamma$-matrices to be real.
We will denote $\rho_{SU(2,2)}$
the space of the positive chirality 
spinor representation of $SO(2,4)$ (the fundamental of $SU(2,2)$)
and $\rho_{SU(4)}$ the space of the positive chirality
spinor representation of $SO(6)$ 
(the fundamental of $SU(4)$). Therefore 
$\Psi_{++}\in \rho_{SU(2,2)}\otimes \rho_{SU(4)}$.

The worldsheet degrees of
freedom are two Majorana-Weyl spinors
$\theta^1(\tau,\sigma)$, $\theta^2(\tau,\sigma)$.
We will parametrize them by a single complex $\Psi_{++}$:
\begin{eqnarray}
\theta^1=(1+\Gamma^A\Gamma^S)\mbox{Re}(\Psi_{++})\nonumber
\\[5pt]
\theta^2=(1+\Gamma^A\Gamma^S)\mbox{Im}(\Psi_{++})
\end{eqnarray}
The $\gamma$-matrices of \cite{MetsaevTseytlin} are related to
the gamma-matrices in the tangent space to $AdS_5\times S^5$:
\begin{equation}
i\gamma^a=\Gamma^a\hat{F}_A,\;\;\;\gamma^{a'}=\Gamma^{a'}\hat{F}_S,\;\;\;
\hat{F}_A\theta^I=\hat{F}_S\theta^I
\end{equation}
Here $\hat{F}_A$ is the product of the five gamma-matrices $\Gamma_a$ tangent
to $AdS_5$ and $\hat{F}_S$ is the product of gamma-matrices tangent to $S^5$. 
Also, for any vector $v$ in the tangent space to $AdS_5\times S^5$ we will 
denote:
$$\hat{v}=\Gamma_{\mu}v^{\mu}$$
Let us introduce a notation:
\begin{equation}\label{ConjugationOfRealSpinor}
	\overline{\theta}=\theta^{T}\Gamma_{-1}\Gamma_0
\end{equation}
where the superindex $T$ denotes the transposition.
In this definition we assume that we have chosen the 
gamma-matrices so that $\Gamma_{-1}$ and $\Gamma_0$
are antisymmetric, and $\Gamma_1,\ldots, \Gamma_{10}$
are symmetric.
The definition (\ref{ConjugationOfRealSpinor}) 
is for a real spinor $\theta$.
For a complex combination $\theta_1+i\theta_2$ we define
\begin{equation}
	\overline{\theta_1+i\theta_2}=
	\overline{\theta}_1-i\overline{\theta}_2
\end{equation}
This notation allows
us to write an explicit formula for the linear map
from the tensor product of two chiral spinor bundles to the
vector bundle:
\begin{equation}
	\theta_1\otimes \theta_2 \mapsto j,\;\;\;\;\;
	j^{\mu}=\overline{\theta}_1\Gamma^A
	\Gamma^{\mu}\theta_2
\end{equation}
We will use the same notation for the conjugate of $\Psi_{++}$:
\begin{equation}
	\overline{\Psi_{++}}=\Psi_{++}^{*T}\Gamma_{-1}\Gamma_0
\end{equation}
where $\Psi_{++}^*$ means the complex conjugate
of $\Psi_{++}$ (notice that $\Psi_{++}$  has 
to be a complex spinor).

The covariant derivative modified by the Ramond-Ramond
field strength is:
\begin{equation}
	{\cal D}_i(\theta^1+i\theta^2)=
	\left[D_i+
	{1\over 4}i(\hat{F}_A-\hat{F}_S)\Gamma_i\right](\theta^1+i\theta^2)
\end{equation}
The main advantage of considering ten-dimensional spinors
as restrictions of twelve-dimensional spinors is a simple
form of the covariant derivative:
\begin{equation}
{\cal D}_i \left[(1+\Gamma^A\Gamma^S)\Psi_{++}\right]=
(1+\Gamma^A\Gamma^S)\partial_{i}\Psi_{++}
\end{equation}
This means that covariantly constant spinors 
correspond to constant $\Psi_{++}$.

In this paper we will restrict ourselves to the study of the
configurations near $\theta^I=0$; we will only keep the terms of
the lowest order in $\theta^I$ (terms quadratic in $\theta^I$ in 
the action).
With this restriction, the kappa-transformations are:
\begin{eqnarray}
	\delta_k \theta^1=\widehat{\partial_+ x}\; k^1\\[5pt]
	\delta_k \theta^2=\widehat{\partial_- x}\; k^2
\end{eqnarray}
and the equations of motion for fermions are:
\begin{eqnarray}
	\widehat{\partial_+x}\;{\cal D}_-\theta^1=0\\[5pt]
	\widehat{\partial_-x}\;{\cal D}_+\theta^2=0
\end{eqnarray}
They imply that there are spinors $\eta^1,\eta^2$ such that
\begin{eqnarray}
{\cal D}_-\theta^1=\widehat{\partial_+x}\;\eta^1\\[5pt]
{\cal D}_+\theta^2=\widehat{\partial_-x}\;\eta^2
\end{eqnarray}
Doing the kappa-transformation with the parameters
$k^1,k^2$ such that $D_- k^1=-\eta^1$ and
$D_+ k^2=-\eta^2$ we are left with
\begin{equation}
{\cal D}_-\theta^1={\cal D}_+\theta^2=0
\end{equation}
There are some
 ``residual'' kappa-transformations which preserve this condition.

For the fast moving string, we choose the coordinates
$\tau,\sigma$ so that $g_{\tau\tau}=\epsilon^2$, 
$g_{\sigma\sigma}=-1$, $g_{\tau\sigma}=0$.
The equations of motion in the ``complex'' notations is:
\begin{equation}
{\cal D}_{\tau}(\theta^1+i\theta^2)-
\epsilon \left({\cal D}_{\sigma}(\theta^1+i\theta^2)\right)^*=0
\end{equation}
In terms of $\Psi_{++}$:
\begin{equation}
\partial_{\tau}\Psi_{++}-
\epsilon\;\Gamma^A\Gamma^S (\partial_{\sigma}\Psi_{++})^*=0
\end{equation}
 What happens when $\epsilon\to 0$?
We have $\partial_{\tau}\Psi_{++}=0$, thus $\Psi_{++}$ is constant
on the light rays forming the null surface. Let us study the
residual kappa-transformations. Let $\partial_{\tau}x_A$ be
the AdS-component of the tangent vector to the null-geodesic,
and $\partial_{\tau}x_S$ be the component in the tangent space
to the sphere. The following kappa-transformation
with the constant parameter $K_{++}$ 
leaves $\Psi_{++}$ constant along the light rays:
\begin{eqnarray}
\delta_K\left[(1+\Gamma^A\Gamma^S)\Psi_{++}\right]&=&
(\widehat{\partial_{\tau}x_A}+\widehat{\partial_{\tau}x_S})
(1+\Gamma^A\Gamma^S)\Gamma^A K_{++}=\nonumber\\[5pt]
&=&(1+\Gamma^A\Gamma^S)\;
(\widehat{\partial_{\tau}x_A}\Gamma^A+
\widehat{\partial_{\tau}x_S}\Gamma^S)K_{++}\nonumber
\end{eqnarray}
In other words,
\begin{equation}\label{Residual}
\delta_K\Psi_{++}=(\widehat{\partial_{\tau}x_A}\Gamma^A+
\widehat{\partial_{\tau}x_S}\Gamma^S)K_{++}
\end{equation}
The right hand side is constant on the light ray,
because $\widehat{\partial_{\tau}x_A}\Gamma^A$ is the
rotation in the equatorial plane of $AdS_5$ and
$\widehat{\partial_{\tau}x_S}\Gamma^S$ is the rotation
in the equatorial plane of $S^5$. 
These generators of rotations are very useful.
In the spinor language the equator of 
$AdS_5$ corresponds to the 2-plane $L_A\subset \rho_{SU(2,2)}$ which is 
defined as the subspace on which the rotation in the
equatorial plane acts with the eigenvalue $+i$.
And the equatorial plane of $S^5$ corresponds to the
2-plane $L_S\subset\rho_{SU(4)}$. We have a decomposition:
\begin{equation}
\rho_{SU(2,2)}\otimes \rho_{SU(4)}=
(L_A\otimes L_S)\oplus (L_A^{\perp}\otimes S_S^{\perp})
\oplus (L_A\otimes S_S^{\perp})\oplus (L_A^{\perp}\otimes L_S)
\end{equation}
The kappa-transformation (\ref{Residual}) then implies that
the component of $\Psi_{++}$ which belongs to 
$(L_A\otimes L_S)\oplus (L_A^{\perp}\otimes L_S^{\perp})$
is a pure gauge; it can be gauged away.
This means that we can choose $\Psi_{++}$ to satisfy
\begin{equation}\label{PsiCondition}
(\widehat{\partial_{\tau}x_A}\Gamma^A+
\widehat{\partial_{\tau}x_S}\Gamma^S)\Psi_{++}=0
\end{equation}
or equivalently
$(\widehat{\partial_{\tau}x_A}-
 \widehat{\partial_{\tau}x_S})\theta^I=0$.
In other words $\Psi_{++}$ can be brought into the form
\begin{equation}\label{GaugeChoiceForPsi}
\Psi_{++}=\widetilde{\eta}\otimes\phi +
\eta^*\otimes\widetilde{\phi}^*
\end{equation}
where $\phi,\widetilde{\phi}^*$ are in the fundamental
representation of $SU(4)$ and $\eta^*,\widetilde{\eta}$
are in the fundamental representation of $SU(2,2)$ and
$\phi\in L_S$, $\widetilde{\phi}^*\in L_S^{\perp}$, 
$\eta^*\in L_A$, $\widetilde{\eta}\in L_A^{\perp}$.
Let us introduce an orthonormal basis in 
$\rho_{SU(2,2)}\oplus \rho_{SU(4)}^*$,
according to the decomposition $L_A\oplus L_A^{\perp}\oplus 
L_S^*\oplus L_S^{*\perp}$. In this basis, the antihermitean matrix
\begin{equation}
\left[
\begin{array}{cccc} 
0 & 0 & 0 & \eta^*\otimes\widetilde{\phi}^* \\
0 & 0 & \widetilde{\eta}\otimes\phi & 0 \\
0 & -\phi^*\otimes\widetilde{\eta}^*           & 0 \\
-\widetilde{\phi}\otimes\eta & 0 & 0 & 0
\end{array}
\right]
\end{equation}
defines an infinitesimal  variation of the $4$-dimensional
plane $L_A\oplus L_S^*$ 
in the $8$-dimensional space $\rho_{SU(2,2)}\oplus \rho_{SU(4)}^*$,
such that the variation of $L_A$ goes outside 
$\rho_{SU(2,2)}$ into $\rho^*_{SU(4)}$, and the variation of $L_S^*$
goes into $\rho_{SU(2,2)}$. The precise definition of
this infinitesimal variation goes as follows. Take four  complex linearly
independent vectors $e_1,e_2,e_3,e_4$ in $L_A\oplus L_S^*$,
so that $L_A\oplus L_S^*$ is generated by $e_1,e_2,e_3,e_4$. 
Given $\Psi_{++}$ of the
form (\ref{GaugeChoiceForPsi}) we define the variation 
\begin{equation}\label{VariationOfBasis}
\delta e_i = (\phi,e_i)\tilde{\eta} - (\eta, e_i)\tilde{\phi}
\end{equation}
The plane generated by the four vectors
$e_1+\delta e_1, e_2+\delta e_2, e_3+\delta e_3, e_4+\delta e_4$
is the infinitesimal variation of $L_A\oplus L_S^*$.

\begin{figure}
\begin{center} 
\epsfxsize=4in {\epsfbox{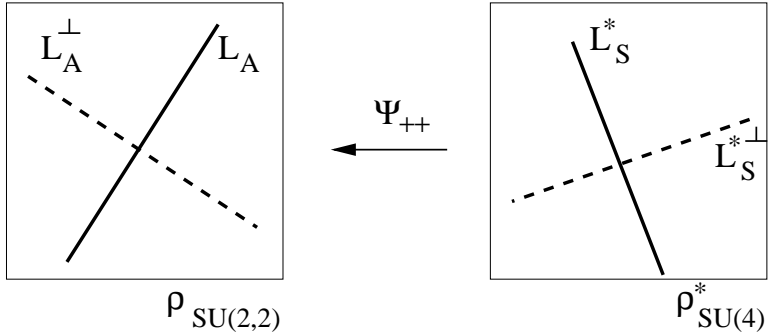}} 
\caption{Eq. (\ref{GaugeChoiceForPsi}) shows that 
$\Psi_{++}\in \rho_{SU(2,2)}\otimes \rho_{SU(4)}$
defines a linear map
$\Psi_{++}:\;\rho_{SU(4)}^*\to \rho_{SU(2,2)}$ such
that $L_S^*$ goes into  $L_A^{\perp}$ and 
$L_S^{*\perp}$ goes into $L_A$. Therefore 
$(\Psi_{++},\Psi_{++}^*)$ defines a linear automorphism of 
$\rho_{SU(2,2)}\oplus \rho_{SU(4)}^*$
and theorefore an infinitesimal
variation of the plane $L_A\oplus L_S^*\subset \rho_{SU(2,2)}\oplus 
\rho_{SU(4)}^*$.}
\end{center} 
\end{figure}

We have the following picture. The null-surface is a collection of the
light rays. Each light ray defines an equator in $AdS_5$ or equivalently
 a 2-plane $L_A\subset \rho_{SU(2,2)}$, and an equator
in $S^5$ or equivalently a 2-plane $L_S^*\subset \rho^*_{SU(4)}$.
 Turning on $\theta^I$ corresponds
to the deformation of $L_A\oplus L_S^*$ inside 
$\rho_{SU(2,2)}\oplus \rho^*_{SU(4)}$, so that $L_A$ is not entirely
inside of $\rho_{SU(2,2)}$ and $L_S^*$ is not entirely in $\rho^*_{SU(4)}$.
In other words, while a ``purely bosonic'' null ray parametrizes
a pair of 2-planes 
$$(L_A\subset \rho_{SU(2,2)},\; L_S^*\subset\rho^*_{SU(4)})$$
a null ray with $\theta^I$ parametrizes  a 4-plane
$$L_A\oplus L_S^*\subset \rho_{SU(2,2)}\oplus \rho^*_{SU(4)}$$
without the constraints that $L_A$ belongs to $\rho_{SU(2,2)}$
or $L^*_S$ belongs to $\rho^*_{SU(4)}$.
But we also have to take into account that $\theta^I$ are
odd coordinates. This can be done by declaring $\rho_{SU(2,2)}$ 
an ``even'' space and $\rho^*_{SU(4)}$ an ``odd'' space. The total space
is now $\rho_{SU(2,2)}\oplus \Pi\rho^*_{SU(4)}$ where
$\Pi$ means that the vector space is considered odd. 
And our 4-plane $L_A\oplus L_S^*$ is actually $L_A\oplus \Pi L_S^*$.
We are embedding the complex space of dimension $(2|2)$ in
the complex space of dimension $(4|4)$:
\begin{equation}
L_A\oplus \Pi L_S^*\;\;\subset \;\;\rho_{SU(2,2)}\oplus \Pi \rho^*_{SU(4)}
\end{equation}
This means that turning on the fermionic degrees of freedom
replaces the product of two ordinary Grassmanians with
the super-Grassmanian:
\begin{equation}\label{SuperGrassmanian}
	{SU(2,2)\over S(U(2)\times U(2))}\times {SU(4)\over S(U(2)\times U(2))}
	\to {U(2,2|4)\over U(2|2)\times U(2|2)}
\end{equation}
Since $L_A\oplus \Pi L_S^*$ is a complex space, the super-Grassmanian
is a complex supermanifold. 
Eq. (\ref{VariationOfBasis}) implies that $\tilde{\eta}\otimes\phi$
and $\tilde{\phi}\otimes\eta$ are holomorphic coordinates.
Therefore the complex structure acts on $\Psi_{++}$
as follows:
\begin{equation}\label{ComplexStructure}
	I.\Psi_{++}=I.(\widetilde{\eta}\otimes\phi +
\eta^*\otimes\widetilde{\phi}^*)=(i\widetilde{\eta}\otimes\phi-
i\eta^*\otimes\widetilde{\phi}^*)=
\widehat{\partial_{\tau}x_A}\Gamma^A\;\Psi_{++}
\end{equation}
Super-Grassmanians are discussed in Chapter 4 of \cite{Manin}.

\subsection{The Hamiltonian system.}
The space of parametrized null surfaces
comes with a natural Hamiltonian dynamics. The Hamiltonian as a functional
of a contour in the coset space is equal to the action of the
particle, which has this contour as a trajectory. The symplectic
form is the Kahler form on the coset integrated over the
contour. (It is obtained as a limit of the natural symplectic
form on the space of the classical string worldsheets.)
With the fermionic degrees of freedom turned on,
the Hamiltonian is the action of the particle moving on
the super-Grassmanian (\ref{SuperGrassmanian}).
The metric on this space is fixed by the supersymmetry, and the
symplectic form follows from the symplectic form of the
string worldsheet theory. The action of the string worldsheet
theory is:
\begin{equation}\label{StringWorldsheet}
	{1\over\epsilon}\int d\tau d\sigma\left\{
	 (\partial_{\tau}x)^2-\epsilon^2(\partial_{\sigma}x)^2
	 +(\overline{\theta^1}, \Gamma^A 
	 \widehat{\partial_+ x} \; D_-\theta^1)
	 +(\overline{\theta^2}, \Gamma^A 
	 \widehat{\partial_- x} \; D_+\theta^2)\right\}
\end{equation}
To compute the symplectic form at the leading order in
$1\over \epsilon$ we substitute 
for $x$ and $\theta$ a parametrized null-surface.
We get for the symplectic form\footnote{The relative sign of
 $|dY|^2$ and $|dZ|^2$ is different from what we had in (\ref{Delta})
 because in this section 
 we are using the ``mostly plus'' convention for the metric
 following \cite{MetsaevTseytlin}. With this convention
 $|dY|^2=-|dY_{-1}|^2-|dY_0|^2+\sum_{i=1}^4 |dY_i|^2$ and
 $|dZ|^2=\sum_{I=1}^6 |dZ_I|^2$.}:
\begin{eqnarray}\label{SuperSymplectic}
	&&\hspace{-20pt}\omega={1\over\epsilon}d\phi\wedge dE +
	\nonumber\\[5pt] &&
	\hspace{-20pt}
+{1\over\epsilon}
\int d\sigma\;\left[(\overline{dY}\wedge dY)+(\overline{dZ}\wedge dZ)+
\left(d\left(\overline{\Psi_{++}}\;
[\widehat{\overline{Y}},\widehat{Y}]\right)\wedge d\Psi_{++}\right)\right] 
\end{eqnarray}
where $\phi$ is the relative phase of $Y$ and $Z$, and $E$ is
its conjugate variable (see \cite{SE}). This expression is written
in the gauge $(\overline{Y},\partial_{\sigma}Y)=
(\overline{Z},\partial_{\sigma}Z)=\mbox{const}$ and
with the condition (\ref{PsiCondition}).

A surprising feature of the Hamiltonian is that it
contains a fermionic bilinear with two derivatives. Therefore the
evolution equation for the fermion should be of the form 
\begin{equation}\label{Schroedinger}
\partial_t \Psi_{++}=\partial^2_{\sigma}\Psi_{++}
\end{equation}
In the Metsaev-Tseytlin action the fermion derivatives entered only in
the combinations of the form $\theta d\theta$. Therefore one would expect
equations of motion of the form 
$\partial_t\Psi_{++}=\partial_{\sigma}\Psi_{++}$
rather than (\ref{Schroedinger}). But it turns out that after taking the
average of the equation of motion over the period the first derivative
$\partial_{\sigma}\Psi_{++}$ is replaced by the second derivative. 
Indeed, let us consider the equation for fermions:
\begin{equation}
	\partial_{\tau}\Psi_{++}-
	\epsilon \;\Gamma^{A}\Gamma^{S}
	(\partial_{\sigma}\Psi_{++})^*=0
\end{equation}
This equations looks suspicious, because the $\tau$-derivative of
$\Psi_{++}$ is of the order $\epsilon$ rather than $\epsilon^2$.
The slow evolution should be on the time scale
$\Delta\tau\sim {1\over\epsilon^2}$, not $1\over\epsilon$.
But remember that we have to take the average over the period. 
If we neglect the terms
of the order $\epsilon^2$ and higher, we will get
\begin{equation}
	\Psi(\tau+2\pi,\sigma)-\Psi(\tau,\sigma)=
	-2\pi\epsilon\;\langle \Gamma^{A}\Gamma^{S} \rangle
	(\partial_{\sigma}\Psi_{++})^*
\end{equation}
Here $\langle \Gamma^{A}\Gamma^{S} \rangle$ means the average
of $\Gamma^{A}\Gamma^{S}$ over the period:
\begin{equation}
\langle \Gamma^{A}\Gamma^{S} \rangle=
{1\over 2}\left( \Gamma^{A}\Gamma^{S}+
\widehat{\partial_{\tau}x}_A
\widehat{\partial_{\tau}x}_S\right)
\end{equation}
But the image of this operator is a kappa-symmetry
(\ref{Residual}):
$$
\left( \Gamma^{A}\Gamma^{S}+
\widehat{\partial_{\tau}x}_A
\widehat{\partial_{\tau}x}_S\right)
(\partial_{\sigma}\Psi_{++})^*=
(\widehat{\partial_{\tau}x_A}\Gamma^A+
\widehat{\partial_{\tau}x_S}\Gamma^S)\;
\widehat{\partial_{\tau}x}_A \Gamma^S(\partial_{\sigma}\Psi_{++})^*
$$
Therefore the slow evolution of $\Psi_{++}$ in the order
$\epsilon$ is trivial.
Let us compute the order $\epsilon^2$. To simplify the calculations,
we will assume that $\partial_{\sigma}x=0$. The variation
of $\Psi$ over the period is:
\begin{eqnarray}
&&\Psi(\tau+2\pi,\sigma)-\Psi(\tau,\sigma)=\nonumber\\[5pt]
&&=-\epsilon^2\int_0^{2\pi}
	d\tau'\int_0^{\tau}d\tau'\;
        \Gamma^A(\tau)\Gamma^S(\tau)
	\Gamma^A(\tau')\Gamma^S(\tau')\;\;\partial^2_{\sigma}\Psi_{++}
\nonumber
\end{eqnarray}
Notice that 
\begin{eqnarray}
&&\Gamma^A(\tau)\Gamma^A(\tau')=-\cos(\tau-\tau')
+\sin(\tau-\tau')\;\widehat{\partial_{\tau}x}_A\Gamma^A\nonumber\\[5pt]
&&\Gamma^S(\tau)\Gamma^S(\tau')=\cos(\tau-\tau')+
\sin(\tau-\tau')\;\widehat{\partial_{\tau}x}_S\Gamma^S\nonumber
\end{eqnarray}
After taking the integrals we get:
\begin{eqnarray}
&&\Psi_{++}(\tau+2\pi,\sigma)-\Psi_{++}(\tau,\sigma)=\nonumber\\[5pt]
&&=\epsilon^2 \left[\pi^2
(1-\widehat{\partial_{\tau}x}_A\Gamma^A 
\widehat{\partial_{\tau}x}_S\Gamma^S)
-{\pi\over 2}(\widehat{\partial_{\tau}x}_A\Gamma^A
-\widehat{\partial_{\tau}x}_S\Gamma^S)\right]
\partial^2_{\sigma}\Psi_{++}\nonumber
\end{eqnarray}
The first term in the square brackets is again a kappa-symmetry 
(\ref{Residual}). The second term is
an operator constant on the light ray, multiplying 
the second derivative of $\Psi_{++}$. 
Now we can introduce the slow time $t=\epsilon^2\tau$ and
write down the equation for the slow evolution:
\begin{equation}
\partial_t\Psi_{++}={1\over 2}\widehat{\partial_{\tau}x}_A\Gamma^A
\;\partial_{\sigma}^2\Psi_{++}
\end{equation}
The structure of this 
equation implies that the Hamiltonian of the slow evolution is of the form
\begin{equation}\label{SuperHamiltonian}
	H=\int d\sigma\;\left[
	(\partial_{\sigma}\overline{Y},\partial_{\sigma}Y)+
	(\partial_{\sigma}\overline{Z}, \partial_{\sigma}Z) + 
	(\partial_{\sigma}\overline{\Psi}_{++},\partial_{\sigma}\Psi_{++})
	\right]
\end{equation}
We should stress that this expression for the Hamiltonian
is valid only in the quadratic order in $\Psi_{++}$. 
The difference in the structure of the fermionic 
term in (\ref{SuperSymplectic}) and  (\ref{SuperHamiltonian})
is related to the action of the complex structure on
$\Psi_{++}$, see Eq. (\ref{ComplexStructure}). 

The terms of the higher order in $\theta$ 
should be fixed by the 
supersymmetry. 
In fact, the quadratic terms are also fixed by the superconformal
symmetry and locality.
 Still, we think it is a nontrivial fact that
the moduli space of null-surfaces in AdS times a sphere
has a natural structure of the Hamiltonian system, and
that the degrees of freedom are roughly the same as
needed to parametrize the single trace operators composed
of the large number of elementary fields. And if it
is possible to understand the higher loop dynamics 
along the lines of \cite{AS,KMMZ,KRT}, it is very unlikely
that it would be also fixed by the superconformal symmetry.

\section*{Acknowledgments}
I would like to thank I.~Bars, J.~Gomis, A.~Kapustin, J.H.~Schwarz 
and E.~Witten for discussions. 
This research was supported by the Sherman Fairchild 
Fellowship and in part
by the RFBR Grant No.  03-02-17373 and in part by the 
Russian Grant for the support of the scientific schools
NSh-1999.2003.2.

\end{document}